\pdfoutput=1

\documentclass[preprint,12pt,showkeys]{revtex4}
\usepackage{amsmath}
\usepackage{latexsym}
\usepackage{amssymb}
\usepackage{graphics,epstopdf}
\usepackage{amsmath,amssymb,amsthm}
\usepackage{slashed}
\usepackage[colorlinks=true, citecolor=blue, urlcolor=blue ]{hyperref}

\begin{document}
\title{Modified Hamiltonian formalism for Regge Teitelboim Cosmology}
\author{Pinaki Patra}

\email{monk.ju@gmail.com}
\affiliation{Department of Physics, University of Kalyani, India-741235}
\author{Md. Raju}

\email{quantumraju@rediffmail.com}
\affiliation{Department of Physics, University of Kalyani, India-741235}
\author{Gargi Manna}

\email{gargi20manna@gmail.com}
\affiliation{Department of Physics, University of Kalyani, India-741235}
\author{Jyoti Prasad Saha}

\email{jyotiprasadsaha@gmail.com}
\affiliation{Department of Physics, University of Kalyani, India-741235}
\begin{abstract}
The Ostrogradski approach for the Hamiltonian formalism of higher derivative theory is not satisfactory because
 of the reason that the Lagrangian cannot be viewed as a function on the tangent bundle to coordinate manifold. 
In this article, we have used an alternative approach which leads directly to the Lagrangian which, being a function on the tangent manifold,
 gives correct equation of motion; no new coordinate variables need to be added. This approach can be directly used to the singular
 (in Ostrogradski sense) Lagrangian. We have used this method for the Regge Teitelboim (RT) minisuperspace cosmological model. 
We have obtained the Hamiltonian of the dynamical equation of the scale factor of RT model.
\end{abstract}


\maketitle
\section{Introduction}
It is fairly  well known that, adding higher derivatives term in Lagrangian may improve the theory in some respects, like ultraviolet behavior
 \cite{Thiring, Pais}, gravity renormalization \cite{Stelle} or even making modified gravity as asymptotically free \cite{Fradkin}).
 Also, higher-derivative Lagrangians appear to be a useful tool to describe some interesting models, like relativistic particles with rigidity,
 curvature, and torsion \cite{Yu,pis,nes,ply,ply1,ban}. \\
The starting of the study of higher derivative theory was started long back. Classical dynamics of a test particle's motion with
higher-order time derivatives of the coordinates was first described in 1850 by Ostrogradski \cite{Ostrogradski} and is known as Ostrogradski's Formalism
which has been extensively studied by several authors for its wide applicability \cite{pod1,pod2,pod3,ili,eli,neu,noj,chu,carr,ani,wood,cor,and,ber,gama}. \\
An interesting occurrence of higher derivative terms in the action appears in general relativity. In some cases, such
terms are isolated as surface terms and dropped. However, in the case of gravity, the surface term is never ignorable,
e.g., the requirement of the Gibbons-Hawking term in the action. This is more so in the brane world scenario where
the universe is viewed as a hyper-surface immersed in a
bulk. A classic model is due to Regge and Teitelboim (RT) \cite{RT}, where gravitation is described as the world volume
swept out by the motion of a three-dimensional brane in a higher dimensional Minkowski space-time. Hamiltonian
analysis of the model and its quantization was further explored in \cite{RT1, RT2, RT3}. Unlike the Einstein gravity, in the RT
model the independent fields are the embedding functions rather than the metric. In the RT model second derivatives
of the fields appear in the action, and like general relativity these higher derivative terms may be clubbed in a surface
term. In the usual formulation this surface term is dropped \cite{RT2}, thereby reducing the original model to a first-order
theory. However, this makes the Hamiltonian formulation of the model problematic \cite{RT2}. These problems are bypassed
by introducing an auxiliary field \cite{RT2}. On the other hand, recently it has been pointed out that no such auxiliary field
is needed if one includes the surface term in the RT model containing higher derivative terms \cite{RT3}.
 Obviously, therefore, the Hamiltonian formulation of this model is far from closed. The analysis of the RT model in the ambit of higher derivative
theory \cite{RT3} was done from the Ostrogradsky approach and this work was based on the minisuperspace model following
from the RT theory. The minisuperspace model carries the re-parametrization invariance of the original RT gravity
which appears as gauge invariance in the Hamiltonian analysis. \\
However, the main disadvantage of the Ostrogradski approach is that the Hamiltonian, being a linear function of some momenta, 
is necessarily unbounded from below. In general, this cannot be cured by trying to devise an alternative canonical formalism. 
In fact, any Hamiltonian is an integral of motion, while it is by far not obvious that a generic system described by higher derivative
Lagrangians possesses globally defined integrals of motion, except the one related to time translation invariance. Moreover,
 the instability of the Ostrogradski Hamiltonian is not related to finite domains in phase space, which implies that it
 will survive in the standard quantization procedure (i.e., it cannot be cured by the uncertainty principle). 
The Ostrogradski approach also has some other disadvantages. There is no straightforward transition from the Lagrangian to the Hamiltonian formalism.
\\
Recently, Andrzejewski et. al. \cite{Andrzejewski} have proposed a modified formalism for the higher derivative theory which can cure
some of the drawbacks of Ostrogradski formalism. Basic idea is the same as that of Ostrogradski. But, the advantage of this approach 
is that the Legendre transformation can be performed in a straightforward way. Though, the Hamiltonian of the modified formalism
 is directly connected to the Hamiltonian obtained in Ostrogradski formalism through a canonical transformation.
\\
In this article we have used the modified formalism proposed by Andrzejewski et. al. \cite{Andrzejewski} for the Regge Teitelboim (RT)
 minisuperspace cosmological model. 
\section{Regge Teitelboim Cosmological Model}
The Regge Teitelboin Cosmological model has been studied in \cite{rabin}. We include this section  for the completeness of our article and we used there notation.
The RT model considers a $d-dimensional$ brane $\Sigma$ which evolves in an $N-dimensional$ bulk space-time with fixed
Minkowski metric $\eta_{\mu\nu}$. The world volume swept out by the brane is a $d+1$-dimensional manifold $m$ defined by the
embedding $x^\mu =X^\mu \zeta(a)$; where $x^\mu$ are the local coordinates
of the background space-time and $\zeta (a)$ are local coordinates for $m$. The theory is given by the action functional
\begin{equation}
S[X]=\int_m d^{d+1}\zeta \sqrt{-g} (\frac{\beta}{2}\mathcal{R}- \Lambda)
\end{equation}
where $\beta$ has the dimension $[L]^{1-d}$ and $g$ is the determinant of the induced metric $g_{ab}$ . $\Lambda$ denotes the cosmological
constant and $R$ is the Ricci scalar. As has been already stated above, we will be confined to the minisuperspace
cosmological model following from the RT model. \\
The standard procedure in cosmology is to assume that on the large scale the universe is homogeneous and isotropic.
These special symmetries enable the four-dimensional world volume representing the evolving universe to be embedded in a five-dimensional Minkowski spacetime,
\begin{equation}
 ds^2 = -dt^2 + da^2+ a^2 d\Omega_3^2
\end{equation}
where $d\Omega_3^2$ is the metric for unit 3 sphere. To ensure the FRW case, we take the following parametric representation for the brane,
\begin{equation}
 x^\mu = X^\mu (\zeta ^a)= (t(\tau), a(\tau), \chi, \theta, \phi)
\end{equation}
where $a(\tau)$ is known as the scale factor. \\
After ADM  decomposition \cite{RT4, RT5} with space-like unit normals ($N=\sqrt{\dot{t}^2-\dot{a}^2}$ is the lapse function),
\begin{equation}
 n_\mu =\frac{1}{N}(-\dot{a},\dot{t},0,0,0)
\end{equation}
the induced metric on the world volume is given by
\begin{equation}
 ds^2 = -N^2 d\tau ^2 + a^2 d\Omega_3^2
\end{equation}
Now, one can compute the Ricci scalar which is given by
\begin{equation}
 \mathcal{R}= \frac{6\dot{t}}{a^2 N^4} (a\ddot{a}\dot{t}-a\dot{a}\ddot{t}+ N^2 \dot{t})
\end{equation}
With these functions we can easily construct the Lagrangian density as
\begin{equation}
\mathcal{L} = \sqrt{-g} (\frac{\beta}{2}\mathcal{R}- \Lambda)
\end{equation}

The Lagrangian in terms of arbitrary parameter $\tau$ can be
written as \cite{RT3}
\begin{equation}
\mathcal{L}(a,\dot{a},\ddot{a},\dot{t},\ddot{t})= \frac{a\dot{t}}{N^3}
(a\ddot{a}\dot{t}-a\dot{a}\ddot{t}+N^2\dot{t})-Na^3H^2
\end{equation}
where,
\begin{equation}
{H^2=\frac{\Lambda}{3\beta}}
\end{equation}
$H$ is called the Hubble-parameter. \\

Varying the action with respect to $a(\tau)$, we get the corresponding Euler-Lagrange equation
\begin{equation}
 \frac{d}{d\tau} \left(\frac{\dot{a}}{\dot{t}}\right) = - \frac{N^2 (\dot{t}^2 - 3N^2 a^2 H^2)}{a\dot{t}(3\dot{t}^2- N^2 a^2 H^2)}
\end{equation}

Please note that the Lagrangian contains higher derivative terms of field $a$ . However, we can write it as \cite{RT3}
\begin{equation}
 L= -\frac{a \dot{a}^2}{N} + a N (1-a^2 H^2)+\frac{d}{d\tau} \left(\frac{a^2 \dot{a}}{N}\right)
\end{equation}

If we neglect the boundary term, the resulting Lagrangian becomes the usual first-order one. However the Hamiltonian analysis is facilitated
 if we retain the higher derivative term. Thus our Hamiltonian analysis will proceed
from the above equation containing higher derivative term. Note that the higher-order model was also considered in \cite{RT3},
 where the Hamiltonian analysis was performed following the Ostrogradsky approach. We, on the
contrary, shall follow the equivalent first-order approach of Andrzejewski et. al. \cite{Andrzejewski}.

\section{Hamiltonian formalism for Regge Teitelboim Cosmological Model}
Our concerned Lagrangian as mentioned in the previous section is
\begin{equation}
L(a,\dot{a},\ddot{a},\dot{t},\ddot{t})=\frac{a\dot{t}}{N^3} (a\ddot{a}\dot{t}-a\dot{a}\ddot{t}+N^2\dot{t})-Na^3H^2
\end{equation}
Where,
\begin{equation}
N=\sqrt{\dot{t}-\dot{a}^2}
\end{equation}
and the Hubble parameter which we are considering to be a constant in the present discussion
\begin{equation}
{H^2=\frac{\Lambda}{3\beta}}
\end{equation}

We set for the notational convenience,
\begin{equation}
a=q^{1}_{1} \;,\;\; t=q^{2}_{1}\;,\;\; \dot{a}=\dot{q^{1}_{1}} \;,\;\; \dot{t}=\dot{q^{2}_{1}} \;,\;\; \ddot{a}={q^{1}_{2}} \;, \;\; \ddot{t}={q^{2}_{2}}
\end{equation}
Our Lagrangian is singular in Ostrogradski sense. Because,
\begin{eqnarray}
\mbox{det}(W_{\mu \nu})= \mbox{det} \left(\frac{\partial^{2}L}{\partial \ddot{q}^\mu \partial
\ddot{q}^\nu}\right)= \mbox{det} \left[ {\begin{array}{cc}
 \frac{\partial^2L}{\partial\ddot{a}\partial\ddot{a}} &  \frac{\partial^2L}{\partial\ddot{a}\partial\ddot{t}} \\
 \frac{\partial^2L}{\partial\ddot{t}\partial\ddot{a}} & \frac{\partial^2L}{\partial\ddot{t}\partial\ddot{t}} \end{array}}\right]\\
 =\mbox{det} \left[ {\begin{array}{cc}
 0 & 0 \\ 0 & 0
\end{array}}\right]=0
\end{eqnarray}
Now, we define $F(q^1_1,q^2_1,\dot{q}^1_1,\dot{q}^2_1,q^1_3,q^2_3)$  such that
\begin{eqnarray}
\mbox{det}(\frac{\partial^{2}F}{\partial\dot{q}^\mu_{1} \partial{q}^\nu_{3}})=\left[ \begin{array}{cc}
\frac{\partial^{2} F}{\partial\dot{q}^{1}_{1} \partial q^{1}_{3}} & \frac{\partial^{2}F}
{\partial\dot{q}^{1}_{1}\partial {q}^{2}_{3}} \\ \frac{\partial^{2}F} {\partial\dot{q}^{2}_{1}\partial {q}^{1}_{3}} &
\frac{\partial^{2}F}{\partial\dot{q}^{2}_{1} \partial {q}^{2}_{3}}\end{array}\right] \neq 0
\end{eqnarray}
One possible choice is
\begin{equation}
F(q^1_1,q^2_1,\dot{q}^1_1,\dot{q}^2_1,q^1_3,q^2_3) = \alpha (\dot{q}^1_1q^1_3 + \dot{q}^2_1q^2_3) + \beta (\dot{q}^1_1 q^2_3 + \dot{q}^2_1q^1_3) \;\;
;\;\; \Delta = \alpha^2 - \beta^2 \neq 0
\end{equation}
Now, using the suggestions prescribed in \cite{Andrzejewski} we define  
\begin{eqnarray}
\mathcal{L}(q^1_1,q^2_1, \dot{q}^1_1,\dot{q}^2_1,q^1_2, q^2_2,q^1_3,q^2_3, \dot{q}^1_3,\dot{q}^2_3) = L + \frac{\partial F}{\partial q_1^\mu} \dot{q}_1^\mu 
+ \frac{\partial F}{\partial q_3^\mu} \dot{q}_3^\mu  + \frac{\partial F}{\partial \dot{q}_1^\mu}q_2^\mu \nonumber \\ = \frac{q^1_1\dot{q}^2_1}{N^3}
 (q^1_1q^1_2\dot{q}^2_1-q^1_1\dot{q}^1_1q^2_2+N^2\dot{q}^2_1)  - N(q^1_1)^3H^2+(\alpha\dot{q}^1_1 + \beta\dot{q}^2_1)\dot{q}^1_3 \nonumber \\ +
(\beta\dot{q}^1_1+\alpha\dot{q}^2_1)\dot{q}^2_3+q^1_2 (\alpha{q^1_3}+\beta{q^2_3})+ (\alpha q^2_3+\beta q^1_3)q^2_2
\end{eqnarray}
The conjugate momenta corresponding to the co-ordinates are given by $p_{ij}=\frac{\partial \mathcal{L}}{\partial \dot{q}_i^j}$. In particular
\begin{eqnarray}
p_{11}=\frac{3}{N^5}(q^1_1)^2 \dot{q}^1_1\dot{q}^2_1 (q^1_2\dot{q}^2_1-\dot{q}^1_1 q^2_2)+\frac{q^1_1\dot{q}^2_1}{N^3} (\dot{q}^1_1
\dot{q}^2_1-q^1_1 q^2_2) \nonumber \\ -\frac{H^2}{N}\dot{q}^1_1(q^1_1)^3+\alpha\dot{q}^1_3+\beta\dot{q}^2_3 \\
p_{12}=\frac{3(q^1_1\dot{q}^2_1)^2}{N^5}(\dot{q}^1_1 q^2_2-q^1_2\dot{q}^2_1)+\frac{q^1_1}{N^3}\left(2q^1_1 q^1_2\dot{q}^2_1-q^1_1
\dot{q}^1_1 q^2_2-(\dot{q}^2_1)^3\right)+  \nonumber \\ \frac{q^1_1}{N}\left(2\dot{q}^2_1-H^2(q^1_1)^2 \dot{q}^2_1\right)+(\alpha\dot{q}^2_3+\beta \dot{q}^1_3) \\
p_{21}=0\\
p_{22}=0 \\
p_{31}=\frac{\partial F}{\partial {q}^1_3}=\alpha\dot{q}^1_1+\beta\dot{q}^2_1 \\
p_{32}=\frac{\partial F}{\partial q^2_3}=\beta\dot{q}^1_1+\alpha\dot{q}^2_1
\end{eqnarray}
The equations (23) and (24) provide primary constraint. Eq. (18) enable us to solve $\dot{q}$'s in terms of momenta. These are explicitly given by
\begin{eqnarray}
\dot{q}^1_1=\frac{1}{\Delta} P_{31} \\
\dot{q}^2_1=\frac{1}{\Delta} P_{32} \\
\dot{q}_3^1 = \frac{\alpha\varrho_1 -\beta\varrho_2}{\Delta} \\
\dot{q}_3^2 = \frac{\alpha\varrho_2 -\beta\varrho_1}{\Delta}
\end{eqnarray}
Where,
\begin{eqnarray}
P_{31} = \alpha p_{31} - \beta p_{32} \\
P_{32}=(\alpha p_{32} - \beta p_{31}) \\
\varrho_1 = p_{11} + \frac{1}{N \Delta} H^2 (q_1^1)^3 P_{31} - \frac{q_1^1}{N^3 \Delta} P_{32} [\frac{1}{\Delta^2} P_{31} 
 P_{32} - q_1^1 q_2^2] \nonumber \\ - \frac{3 (q_1^1)^2}{N^5 \Delta^3} P_{31} P_{32} [q_2^1 P_{32} - q_2^2 P_{31}] \\
\varrho_2 = p_{12} + \frac{(q_1^1 P_{32})^2}{\Delta} (q_2^1 P_{32} - q_2^2 P_{31}) + \frac{q_1^1}{N^3 \Delta} [q_1^1 P_{31}q_2^2 
+ \frac{1}{\Delta^2} P_{32}^3 \nonumber \\ - 2 q_1^1 q_2^1 P_{32}]  \frac{q_1^1}{N\Delta}  [ H^2 (q_1^1)^2 - 2 ] P_{32}
\end{eqnarray}
Therefore, $N$ reduces to
\begin{equation}
N = \sqrt{\frac{(p_{32})^2 - (p_{31})^2 }{\Delta}}
\end{equation}
Now, the Dirac Hamiltonian ($\mathcal{H}$) is given by 
\begin{equation}
\mathcal{H} = p_{1\mu}\dot{q}_1^\mu - L - \frac{\partial F} {\partial q^\mu_1}\dot{q}^\mu_1 - \frac{\partial F}{\partial \dot{q}^\mu_1}q^\mu_2  + c^\mu p_{2\mu}
\end{equation}
Where, $c^\mu$ are two  Lagrange multipliers enforcing the "primary" constraints
\begin{equation}
\Phi_{1\mu} \equiv p_{2\mu} \approx 0
\end{equation}
The Hamiltonian explicitly reads
\begin{eqnarray}
 \mathcal{H}= \frac{1}{\Delta} (p_{11}P_{31} + p_{12}P_{32}) - \frac{1}{N \Delta^2} q_1^1 (P_{32})^2 - \frac{1}{N^3 \Delta^2} [(q_1^1)^2 q_2^1 (P_{32})^2 
- \nonumber \\ (q_1^1)^2 q_2^2 P_{31} P_{32}] - \alpha (q_3^1 q_2^1 + q_3^2 q_2^2) - 
\beta (q_3^2 q_2^1 + q_3^1 q_2^2) + \nonumber \\ H^2 N (q_1^1)^3 + c^1 p_{21} 
+ c^2 p_{22}
\end{eqnarray}
And the Hamilton's equations of motion are given by
\begin{eqnarray}
\dot{q}_1^1 = \frac{1}{\Delta} P_{31} \\
\dot{p}_{11} = \frac{1}{N\Delta^2} (P_{32})^2 + \frac{2q_1^1}{N^3 \Delta^2}(q_2^1 P_{32} - q_2^2 P_{31}) P_{32} - 3 H^2 N (q_1^1)^2 \\
\dot{q}_1^2 = \frac{1}{\Delta} P_{32} \\
\dot{p}_{12} = 0 \\
\dot{q}_2^1 = c^1 \\
 \dot{p}_{21} = \frac{1}{N^3 \Delta^2} (q_1^1 P_{32})^2 + \alpha q_3^1 + \beta q_3^2 \\
 \dot{q}_2^2 = c^2 \\
  \dot{p}_{22} = - \frac{1}{N^3 \Delta^2} (q_1^1 )^2  P_{31}P_{32}+ \alpha q_3^2 + \beta q_3^1 \\
  \dot{q}_3^1 = \frac{1}{\Delta}(\alpha p_{11}-\beta p_{12}) -\frac{1}{N^3 \Delta^3} q_1^1 p_{31} 
(P_{32})^2 + \frac{2\beta}{N\Delta^2} q_1^1 P_{32} - \frac{3p_{31}}{N^5 \Delta^3} (q_1^1)^2 P_{32}
[  \nonumber \\ q_2^1 P_{32} -  q_2^2 P_{31}] -\frac{1}{N^3 \Delta^2} (q_1^1)^2 [\beta q_2^2 P_{31} - 
(\alpha q_2^2 + 2\beta q_2^1)P_{32}] - \nonumber \\ \frac{1}{N\Delta} H^2 p_{31} (q_1^1)^3   \\
  \dot{p}_{31} = \alpha q_2^1 + \beta q_2^2\\
  \dot{q}_3^2 = \frac{1}{\Delta}(\alpha p_{12}-\beta p_{11}) + \frac{1}{N^3 \Delta^3} q_1^1 p_{32}
 (P_{32})^2 - \frac{2\alpha}{N\Delta^2} q_1^1 P_{32} + \frac{3p_{32}}{N^5 \Delta^3} (q_1^1)^2 P_{32} 
[ \nonumber \\ q_2^1 P_{32} -  q_2^2 P_{31}] -\frac{1}{N^3 \Delta^2} (q_1^1)^2 [- \alpha q_2^2 P{31} + 
(2 \alpha q_2^1 + \beta q_2^2)P_{32}] + \frac{1}{N\Delta} H^2 p_{32} (q_1^1)^3 \\
  \dot{p}_{32} = \alpha q_2^2 + \beta q_2^1
\end{eqnarray}
The secondary constraints reads
\begin{equation}
0\approx \Phi_{2\mu} = \frac{\partial L(q_1,\dot{q}_1,q_2)}{\partial q_2^\mu} + \frac{\partial F(q_1,\dot{q}_1,q_3)}{\partial \dot{q}_1^\mu}
\end{equation}
i.e.,
\begin{eqnarray}
0 \approx \Phi_{21} = \frac{1}{N^3} (q_1^1 \dot{q}_1^2)^2 + \alpha q_3^1 + \beta q_3^2 \\
0 \approx \Phi_{22} = -  \frac{1}{N^3} (q_1^1)^2 \dot{q}_1^1 \dot{q}_1^2+ \alpha q_3^2 + \beta q_3^1
\end{eqnarray}
To determine $c^\mu$, usually one uses the stability condition of the secondary constraints
\begin{equation}
0\approx \{ \Phi_{2\mu} , \mathcal{H}\}
\end{equation}
But, for our system under consideration, $W$ (eq. (16), (17) has rank $0$. This implies the existence of $2$ linearly independent null 
vectors; so, one can not obtain the values of $c^\mu$ in this case. However, if we are interested in the dynamical equations for $a$ and $t$, 
we can use the Hamiltonian 
\begin{equation}
\mathcal{H} = p_{1\mu}\dot{q}_1^\mu - L - \frac{\partial F} {\partial q^\mu_1}\dot{q}^\mu_1 - \frac{\partial F}{\partial \dot{q}^\mu_1}q^\mu_2  
\end{equation}
which  is explicitly given by 
\begin{eqnarray}
 \mathcal{H}= \frac{1}{\Delta} (p_{11}P_{31} + p_{12}P_{32}) - \frac{1}{N \Delta^2} q_1^1 (P_{32})^2 - \frac{1}{N^3 \Delta^2} [(q_1^1)^2 q_2^1 (P_{32})^2 
-  (q_1^1)^2 q_2^2 P_{31} P_{32}] \nonumber \\ - \alpha (q_3^1 q_2^1 + q_3^2 q_2^2) - 
\beta (q_3^2 q_2^1 + q_3^1 q_2^2) + H^2 N (q_1^1)^3
\end{eqnarray}
If we promote it to quantization, it can easily be seen that $\mathcal{H}$ is hermitian.
\section{Conclusions}
We have obtained the Hamiltonian structure for the scale factor of the RT model. In the modified formalism used in this article, the Legendre transformation
can be performed in a straightforward way. 
Summarizing, we have found modified  Hamiltonian formulations of RT gravity  which is
equivalent to the Ostrogradski formalism in the sense that, they are related to the latter by a
canonical transformation.  \\
The stability condition of constraint for the modified formalism proposed in  \cite{Andrzejewski}, fails to determine the lagrange multipliers for the 
model discussed in this article. In that sense one can conclude that the modified formalism proposed in \cite{Andrzejewski} is not always superior to the usual 
Ostrogradski formalism used in literature.
\section{Acknowledgement}
We would like to thank the anonymous reviewers for their careful reading of our manuscript and for their valuable comments which made this article in present form. 
Pinaki Patra is grateful to CSIR, Govt. of India for fellowship support, Gargi Manna is grateful to DST, Govt. of India for
DST-INSPIRE scholarship, Jyoti Prasad Saha is grateful to DST-PURSE for financial support.

\end{document}